
%
%

\def\etal{\it et al.\rm}
\def\kms{km s$^{-1}$}

\def\gsim{ \lower .75ex \hbox{$\sim$} \llap{\raise .27ex \hbox{$>$}} }
\def\lsim{ \lower .75ex \hbox{$\sim$} \llap{\raise .27ex \hbox{$<$}} }
\def\pp{\noindent\parshape 2 0truecm 16.0truecm 1.0truecm 15truecm}

\def\spose#1{\hbox to 0pt{#1\hss}}
\def\simlt{\mathrel{\spose{\lower 3pt\hbox{$\mathchar"218$}}
     \raise 2.0pt\hbox{$\mathchar"13C$}}}
\def\simgt{\mathrel{\spose{\lower 3pt\hbox{$\mathchar"218$}}
'     \raise 2.0pt\hbox{$\mathchar"13E$}}}
\hrule height0pt
\magnification=\magstep1
\baselineskip 14pt
\parskip=6pt
\parindent=0pt

\hsize=6.5truein
\vsize=8.5truein


\font\titlefont=cmss17


\centerline{\titlefont THE \ NATURE \ OF \ DARK \ MATTER}

\vskip 1.0truein
\centerline{\titlefont Ben \ Moore}
\vskip 0.3truein

\centerline{\it Department of Astronomy, University of California,
Berkeley, CA 94720, USA}
\vskip 0.7truein

\centerline {\bf ABSTRACT}
\vskip 8pt
\parindent=36pt
\vskip 0.3truecm

Collisionless particles, such as cold dark matter, interact only by gravity
and do not have any associated length scale, therefore the dark halos of
galaxies should have negligible core radii. This expectation has been
supported by numerical experiments of collisionless particles within scale
free and cold dark matter cosmologies. Most dwarf spiral galaxies are almost
completely dark matter dominated, allowing a unique insight into their mass
\--- density profiles which can be approximated by isothermal spheres with
core radii of order 3 \--- 7 kpc. We can therefore make a direct comparison
between these galaxies, and halos which form within the numerical
simulations. This yields a severe discrepancy in that the simulations
predict the density to fall as $\rho(r)\propto r^{-1}$ on the scales where
the data show that $\rho(r) = {\rm const}$: {\it e.g.} the models
overestimate the mass within the central few kpc of the halos by a factor of
four. The formation of the luminous component of galaxies exacerbates this
disparity between theory and observations, since the contraction of the
baryons can significantly increase the central dark matter density.

\vfill\eject

One of the primary motivations for invoking dark matter halos arose when
radio observations of neutral hydrogen (HI) in spiral galaxies at large
galactocentric radii, $r_g$, showed evidence for more mass than was observed
within the visible components$^{\bf 1}$. In general, the gas rotates at an
approximately constant velocity out to the last measured data point$^{\bf
2,3}$, implying that the density of material falls approximately as
$r_g^{-2}$, {\it i.e.} the mass increases linearly with distance. For $r_g$
less than the characteristic optical radius, the dynamics are dominated by
the luminous material, whereas on larger scales a dark matter component is
required to maintain the motion of the observed baryons. The importance of
the dark halo at small radii depends upon the mass to light ratio ($M/L$) of
the baryonic material in the disk; typically, $M/L$ ratios of order unity
are adopted. This would imply that in a large spiral galaxy like our own,
the gravity of the luminous material dominates over that of the dark matter
to distances of order $\sim 10$ kpc.

Determining the structure of dark halos from the observations is difficult,
and only in a handful of galaxies is this value tightly constrained due to
the uncertainties in the mass of the disk material and the unknown dark
matter distribution$^{\bf 2}$. Maximum disk models attempt to fit the inner
part of the rotation curve using only the observed bulge and disk components
and adopting high $M/L$ ratios, thus requiring extended dark halos with core
radii as large as 30 kpc$^{\bf 2-5}$. If very low mass to light ratios are
adopted, then the core radii can be significantly reduced, but not
completely. Dwarf spiral galaxies provide excellent probes of the internal
structure of dark halos, since these galaxies are completely dominated by
dark matter on scales larger than a few kpc$^{\bf 6-8}$. Furthermore, most
of the baryonic mass is in the form of HI within the disk, which can be
directly weighed using the observed 21 cm emission.

At present less than a dozen rotation curves have been measured for dwarf
galaxies, however a trend is clearly apparent in that the rotational
velocity of the gas is typically rising over most of the data, many optical
scale lengths into the dark halo, but within the core of the mass
distribution (see Figure 1). For example, the rotation curve of the dwarf
galaxy DDO 154 shown in Figure 1, is dominated by dark matter beyond about
1.5 kpc$^{\bf 8}$. In this system the amount of HI, $2.7\times 10^8M_\odot$,
is 5 times the stellar component and extends over 15 optical scale lengths
into the dark halo which has total mass $\gsim 5\times 10^{9}M_\odot$.
Although dwarf spirals such as the DDO galaxies have luminosities only a few
percent of an $L_*$ galaxy, these galaxies are the most numerous, and a
large fraction of the mass in the Universe is associated with these systems.
For comparison, the rotation curve of NGC 925 is plotted in Figure 1$^{\bf
9}$. This is a normal spiral galaxy, but even this system has a dark halo
with large core radius, $r_c \gsim 7$ kpc.

Any successful cosmological model must be able to reproduce both the
observed large and small scale structures, from galaxy clusters to galaxy
halos. A Universe dominated by collisionless particles, such as axions or
gravitinos which interact only through gravity$^{\bf 10}$, has been
extensively investigated$^{\bf 11-14}$. One of the major successes of the so
called cold dark matter (CDM) models, is the formation of dark halos whose
density profiles fall approximately as $r_g^{-2}$; implying a circular
velocity, $v_c(r_g) = (Gm/r_g)^{1/2}$, which is constant with distance$^{\bf
15,16}$. The extent of the dark halos in these models depends on $\Omega$:
If $\Omega\approx 0.1$, then the halo of our Galaxy would extend to touch
that of Andromeda's at $\approx 300$ kpc.

The inner structure of halos which form within CDM and scale free Universes
dominated by collisionless particles, has recently been re-examined at a
resolution 50 times higher than the original simulations$^{\bf 17,18}$. This
work confirms the original numerical investigations$^{\bf 15,16}$, and
concludes that halos form with nearly flat rotation curves, with small
gradients which depend upon the slope of the initial power spectrum. {\it
Moreover, the density continues to increase to distances less than the
resolution limit of $\sim 1$ kpc, and the core radius, if present, must be
smaller than the resolution scale.} These simulations lend strong support to
the theoretical expectation that the clustering of collisionless dark matter
particles is scale free, and dark halos do not have core radii.

The density profiles of the simulated halos can be fit relatively well using
a Hernquist profile$^{\bf 17}$, where on scales between $10\sim 50$ kpc the
density falls roughly as $r_g^{-2}$, and on small scales as $r_g^{-1}$. The
resulting rotation curve, $v_c(r_g) = (Gm_s r_g)^{1/2} / (r_g + r_s)$, is
parameterised by the effective mass, $m_s$, which can be taken as the mass
within the the virial radius, and $r_s$ is a scale representing the peak
height of the initial density fluctuation$^{\bf 17}$. Dubinski \&
Carlberg$^{\bf 17}$ simulated the formation of a large halo with peak
circular velocity $v_{c,max} \approx 280$ \kms, and obtained values of
$m_s=2.0\times10^{12}M_\odot$ and $r_s=26$ kpc. Warren \etal$^{\bf 18}$
present the rotation curve of a halo from one of their simulations, which
has a peak circular velocity $v_{c,max}\approx 140$ \kms. For this halo the
Hernquist profile with parameters $m_s=2.5\times 10^{11}M_\odot$ and
$r_s=13$ kpc fits very well on scales up to about 20 kpc (although it
appears to fall about 20\% too low on larger scales). These parameters agree
well with those obtained by Dubinski \& Carlberg$^{\bf 17}$, scaled to the
virial mass of the halo, {\it i.e. $m_s\propto v_{c,max}^3$}, and assuming
that $r_s \propto v_{c,max}$. (The rotation curves presented by Warren
\etal$^{\bf 18}$ appear to demonstrate that this scaling holds down to the
smallest halos in their simulations which have amplitude $\approx 60$ \kms.)

Figure 1 shows the predicted rotation curves for several spiral galaxies
obtained using Hernquist profiles with $r_s$ scaled as above, and $m_s$
adjusted so that the curves attain the observed $v_{c,max}$.  The disparity
between the models and the data follows naturally from the nature of
collisionless particles which have no physically associated length scale,
therefore dark matter halos should have zero effective core radii. This
applies to any model in which the dark matter behaves predominantly like
CDM, {\it e.g.} mixed dark matter models$^{\bf 19}$, CDM models with a
cosmological constant$^{\bf 20}$, or baryonic isocurvature fluctuation
models which form compact dark objects very early$^{\bf 21}$.

CDM particles are by definition moving slowly at the present epoch,
therefore unlike hot dark matter particles (HDM), such as light neutrinos,
primordial phase space constraints do not impose a cosmologically
significant scale. Although HDM has been excluded from the dark matter
candidate list for other reasons$^{\bf 22-24}$, phase space constraints do
provide an attractive mechanism to limit the central density of dark matter
particles. An additional phase space constraint might arise through the
merging of halos which tends to populate the available coarse grained phase
space, reducing a pre-existing core until phase space is filled$^{\bf 25}$.
Subsequent merging could possibly reduce the central density of dark matter
particles, but the cosmological simulations discussed above$^{\bf 17,18}$
demonstrate that if this effect occurs, then it is on a scale smaller than
the resolution limit. Numerical effects, such as two body relaxation, also
appear to be unimportant, since identical results were obtained after
increasing the numbers of particles by an order of magnitude, and using
different N-body codes.

The numerical simulations discussed above only follow a dissipationless
component, however gas cools efficiently within dark halos$^{\bf 26}$ and
can modify the halo mass distribution. For example, the adiabatic cooling of
gas which contracts to form the bulge and disk, {\it increases} the central
density of dark matter and would tend to erase any pre-existing core
radii$^{\bf 27,28}$. For a baryonic fraction of 10\%, the final core radius
of a dark halo may have contracted by up to a factor of three after the
baryons have cooled$^{\bf 27}$, thus making the case against exotic dark
matter even stronger. The rotation curves of bright compact spiral galaxies
appear to support this notion, in that these systems have falling rotation
curves due to the large amount of dissipation which has occurred, as
compared with dwarf galaxies or normal spirals which have rising or flat
rotation curves$^{\bf 29}$. Even 100\% efficient energy transfer (in the
opposite sense as discussed above - assuming such a mechanism exists)
between the baryons which could cool within a Hubble time and the dark
matter halos, would be insufficient to imprint a scale the size of the
observed core radii.

Baryonic dark matter has recently achieved renewed popularity, and its
nature is relatively well constrained; low mass stars remain one of the
strongest candidates$^{\bf 30}$. Baryons can interact via electromagnetic
forces as well as gravity and therefore have the potential to introduce a
characteristic scale, perhaps resulting from supernovae driven winds or
cooling flows. The smooth transition between the disk material and halo
material in large spiral galaxies, and the constant ratio of HI surface
density to the dark matter surface density within dwarf galaxies, represent
the well known `disk \--- halo' conspiracies$^{\bf 1,8}$, which are easily
explained if the dark matter is itself baryonic.

We have focussed on the halos of dwarf spiral galaxies whose tractable
density profiles clearly hold important clues as to the nature of dark
matter and the process of galaxy formation. The observations are
inconsistent with the dominant dark matter component being a collisionless,
weakly interacting particle such as invoked by the various CDM scenarios. A
candidate particle with primordial phase space constraints does not exist,
therefore unless the numerical simulations are giving misleading results on
scales larger than 5 times the resolution limit, we are drawn to the
conclusion that the dark matter consists of baryons.

\noindent{\bf Acknowledgements} \ \ I would like to thank Marc Davis for
many useful and interesting discussions.

\vfil\eject

{\noindent{\bf REFERENCES}

\parskip=0pt

\pp 1. \ Faber S.M. \& Gallagher J.S. 1979, {\it Ann.Rev.Ast.Astr.}, {\bf
17}, 135-87.

\pp 2. \ Lake G. \& Feinswog L. 1989, {\it A.J.}, {\bf 98}, 166-79.

\pp 3. \ Begeman K.G. 1989, {\it A.A.}, {\bf 223}, 47-60.

\pp 4. \ Athanassoula E., Bosma A. \& Papaioannou S. 1987, {\it A.A.}, {\bf
179}, 23-40.

\pp 5. \  Kent S.M. 1987, {\it A.J.}, {\bf 93}, 816-32.

\pp 6. \ Jobin M. \& Carignan C. 1990, {\it A.J.}. {\bf 100}, 648-62.

\pp 7. \ Lake G., Schommer R.A. \& van Gorkom J.H. 1990, {\it A.J.}, {\bf
99}, 547-60.

\pp 8. \ Carignan C. \& Beaulieu S. 1989, {\it Ap.J.}, {\bf 347}, 760-70.

\pp 9. \ Wevers B.M.H.R. 1984, {\it Ph.D. thesis}, University of Groningen.

\pp 10. Sarkar S. 1991, {\it NATO A.S.I.}, {\bf 348}, 91-102.

\pp 11. Davis M., Efstathiou G., Frenk C.S. \& White S.D.M. 1985, {\it
Ap.J.}, {\bf 292}, 371-94.

\pp 12. White S.D.M., Frenk C.S., Davis M. \& Efstathiou G. 1987, {\it
Ap.J.}, {\bf 313}, 505-16.

\pp 13. Weinberg D.H. \& Cole S. 1992, {\it M.N.R.A.S.}, {\it 259}, 652-94.

\pp 14. Moore B., Frenk C.S. \& White S.D.M. 1993, {\it M.N.R.A.S.}, {\bf
261}, 827-46.

\pp 15. Frenk C.S., White S.D.M.,  Efstathiou G.P. \& Davis M. 1985, {\it
Nature}, {\bf 317}, 595-7.

\pp 16. Quinn P.J., Salmon J.K \& Zurek W.H., 1986, {\it Nature}, {\bf 322},
329-35.

\pp 17. Dubinski J. \& Carlberg R. 1991, {\it Ap.J.}, {\bf 378}, 496-503.

\pp 18. Warren S.W., Quinn P.J., Salmon J.K. \& Zurek H.W. 1992, {\it
Ap.J.}, {\bf 399}, 405-25.

\pp 19. Davis M., Summers F.J. \& Schlegel D. 1992, {\it Nature}, {\bf 359},
393-96.

\pp 20. Efstathiou G.P., Bond J.R. \& White S.D.M. 1992, {\it M.N.R.A.S.},
{\bf 258}, 1-6P.

\pp 21. Peebles P.J.E. 1987, {\it Nature}, {\bf 327}, 210-11.

\pp 22. White S.D.M., Davis M. \& Frenk C.S. 1984, {\it M.N.R.A.S.}, {\bf
209}, 27-31P.

\pp 23. Spergel D.N. 1990, {\it Nuclear Phys. B.}, {\bf 13}, 66-74.

\pp 24. Gerhard O.E. 1992, {\it Ap.J.Lett.}, {\bf 389}, L9-11.

\pp 25. Pearce F.R., Thomas P.A. \& Couchman H.M.P. 1993, {\it M.N.R.A.S.},
{\bf 264}, 497-508.

\pp 26. Katz N., Hernquist L. \& Weinberg D.H. 1992, {\it Ap.J.Lett.}, {\bf
399}, L109-12.

\pp 27. Blumenthal G., Faber S., Flores G. \& Primack J. 1986, {\it Ap.J.},
{\bf 301}, 27-34.

\pp 28. Carlberg R.G., Lake G. \& Norman C.A. 1986, {\it Ap.J.Lett.}, {\bf
300}, L1-4.

\pp 29. Casertano S. \& van Gorkom J.H. 1991, {\it A.J.}, {\bf 101},
1231-41.

\pp 30. Gerhard O.E. \& Silk J. 1994, {\it Nature} in press.

\vfil\eject

{\noindent{\bf FIGURE 1}}
\vskip 0.2truecm

The observed rotation curves of four spiral galaxies are plotted as a
function of galactocentric distance. The dotted lines show a fit to the
observed rotation curve assuming zero contribution from the baryons, and a
dark halo with an approximately isothermal density profile $\rho(r)\propto
1/(r_c^2+r^2)$, where $r_c$ is the core radius. These `maximum halo' models
fit the dwarf galaxies very well and demonstrate that even if the gravity of
the baryons is ignored, the dark halos must have large core radii. The solid
curves, plotted alongside the data for DDO 154 and NGC 3109, show the
contribution of the dark halo to the rotation curve after including the
contribution of the stellar and HI mass. The core radii of these dark halos
are fairly tightly constrained and show an interesting trend of increasing
size with $v_{c,max}$. (References for the data: DDO 154$^{\bf 8}$, DDO
170$^{\bf 7}$, NGC 3109$^{\bf 6}$, NGC 925$^{\bf 9}$.) The dashed curves
show the rotation curves obtained adopting a Hernquist density profile, and
are drawn fainter on scales below the resolution limit of the simulations.
The effective radius is scaled using $r_s\propto v_{c,max}(\rm observed)$
and the effective mass scaled so that the curve passes through this same
point. (Fixing $m_s$ this way yields curves with amplitudes about 20\% lower
than scaling $m_s \propto v_{c,max}^3$, where $v_{c,max}$ is obtained from
the isothermal halo fit to the data.) The model predictions overestimate the
observed mass within $r_c/2$ by a factor of 4 in each galaxy. The rotation
curve of NGC 3109 implies that for $r<4$ kpc, $\rho(r) \propto r^{0.0}$ to
5\% accuracy, whereas the Hernquist profile predicts $\rho(r) \propto
r^{-1}$. Even if $r_s$ and $m_s$ are left as free parameters, this profile
cannot provide an acceptable fit to the rotation curves of dwarf spirals.

\bye